\documentclass[a4paper,11pt]{article}
\usepackage{graphicx}
\usepackage{subfig}
\usepackage{color}
\usepackage{amsfonts,amssymb,latexsym,amsmath}
\usepackage[section]{placeins}
\usepackage[a4paper,margin=2cm]{geometry}
\usepackage{bm}
\usepackage{array} 
\usepackage{epstopdf}
\usepackage{cite}
\usepackage{float}
\usepackage{hyperref}
\usepackage{array}
\usepackage{adjustbox}
\usepackage{colortbl}
\usepackage{tabularx}
\usepackage{booktabs}
\usepackage{xcolor}

\definecolor{reviewgreen}{RGB}{0,135,0}
\definecolor{revieworange}{RGB}{230,120,0}
\definecolor{reviewblue}{RGB}{0,85,180}

\title{Hybrid Residual Correction of VMC Charmonium Masses with a Screened Funnel Interaction}
\author{
Tarik Akan$^{1}$%
\thanks{
\href{mailto:tarik.akan@bozok.edu.tr}
{tarik.akan@bozok.edu.tr} (Corresponding author)}
,\quad
Metin Yalvac$^{2}$%
\thanks{\href{mailto:metin.yalvac@bozok.edu.tr}
{metin.yalvac@bozok.edu.tr}}
\\[2mm]
$^{1,2}$Physics Department, Yozgat Bozok University,
66100 Yozgat, Turkey
}

\date{\today}

\begin{document}
\maketitle

\begin{abstract}
In this study, we combine residual correction with the physical treatment of charmonium masses
within a Quantum Chromodynamics (QCD) motivated potential-model framework via variational Monte
Carlo (VMC). The aim is not to propose a new charmonium spectrum, since this sector has already been
examined extensively through different potential models. Instead, the main objective is to evaluate
how effectively a Machine Learning (ML) correction can improve a VMC baseline when both stages are
built on the same screened funnel potential. In this workflow, the screened interaction provides the
physical input and determines the underlying mass structure. The proposed run uses the variational
method via VMC and deterministic eigenvalue diagonalization in the process of mass calculation. In
total, ten charmonium states are calculated, among which seven use the experimental reference
masses. The VMC step generates a large set of configurations and local energy estimators that are
fed into the neural network (NN) residual corrector at the sample level, while the corrections and
experimental uncertainties are collected at the state level. It then learns the residual difference
between the raw physics baseline and the internal reference targets. This hybrid procedure reduces
the systematic mass offset of the raw calculation across the studied states. For the experimentally
verified seven states, the correction reduces the MAE from $438.1~\mathrm{MeV}$ to
$24.1~\mathrm{MeV}$, corresponding to a $94.5\%$ reduction. These results show that ML can serve as
a residual-correction layer for potential-model spectroscopy.
\end{abstract}
\section{Introduction}
Charmonium remains a crucial test case for models of heavy-quark dynamics. Its spectrum is
experimentally constrained by the Particle Data Group (PDG) compilations, which provide stable
reference masses for numerical validation
\cite{ParticleDataGroup:2024cfk,patrignani2016review,beringer2012review}. The system also lies in an
energy regime where nonrelativistic intuition remains useful. This feature explains the
long-standing role of potential models in charmonium spectroscopy
\cite{eichten1978charmonium,eichten1980charmonium}. The Cornell interaction established a useful
balance between short-range Coulomb attraction and long-range confinement
\cite{eichten1978charmonium,eichten1980charmonium}. Later relativized approaches extended this
framework by including spin-dependent and kinematic effects \cite{godfrey1985mesons}. Many studies
have refined heavy-quark spectra through modified potentials, improved parameter searches, and
alternative solvers. Flavor-wide spectroscopy constrains heavy-quark interactions across charm and
beauty sectors \cite{eichten1994mesons}. Excited charmonium states provide an additional test of
these interactions because their masses are sensitive to model details \cite{barnes2005_higher}.
Electromagnetic transitions offer another constraint, since higher multipole effects probe the same
wave functions \cite{deng2017_charmonium}. Recent studies have extended this program with
Cornell-type, exponential, quadratic, mixed, and screened potentials
\cite{pathak2022parameterisation,sreelakshmi2022mass,ahmad2025charmonium,akan2025predicting,M2025Mass}.
These works show that potential choice and parameterization strongly affect predicted masses. They
also indicate that another isolated mass fit has limited novelty. The more relevant issue is
computational. In the present work, we use a screened funnel potential that retains the
short-distance Coulomb attraction while replacing the unscreened linear confinement with a softened
long-distance contribution. Even with such a physical baseline, systematic residual structure
may remain. An ML layer can model this residual structure when the baseline preserves the relevant
spectral ordering. For this reason, ML is treated here as a correction to VMC, not as a replacement
for the physical model.

The screened funnel potential is well suited to this purpose because it modifies the confining
interaction at large separation. Color screening has a direct QCD motivation in studies of
heavy-quark bound states at finite temperature and in medium
\cite{karsch1988_color,digal2001_quarkonium}. It is also relevant for excited levels, which probe
larger distances than low-lying states \cite{barnes2005_higher,deng2017_charmonium}. A screened
interaction therefore provides a demanding baseline for residual learning. It contains enough
physical structure to encode short-distance attraction and confinement, while still allowing
systematic numerical deviations to remain. Cornell-based studies already show that the parameter
space of heavy-quark potentials is nontrivial \cite{pathak2022parameterisation,solomko2023cornell}.
The present work follows this logic with a different emphasis. We optimize the screened funnel
parameters for mass resolution within the adopted model and then examine how much of the remaining
discrepancy can be recovered through ML residual correction.

VMC provides a natural physics baseline because it evaluates quantum expectation values
through stochastic sampling. Monte Carlo methods have long been used in nonrelativistic QCD and
heavy-hadron calculations \cite{assi2023baryons,assi2024tetraquarkssufficientlyunequalmassheavy}.
Stochastic reconfiguration also illustrates how Monte Carlo optimization can improve quantum
variational calculations \cite{sorella1998green}. 
These approaches are valuable because they preserve a direct connection between wave functions, Hamiltonians, 
and observables. At the same time, incomplete trial spaces, imperfect parameter choices, or limitations 
of the potential form can generate structured residual errors. Such errors are not merely random fluctuations. 
They may contain learnable information about missing corrections or solver bias. This makes them suitable
targets for supervised ML. 
ML is trained to correct the remaining mass residuals.

ML has become an important practical tool for representation and optimization in quantum many-body
physics. NN wave functions, for example, have shown that artificial neural networks can
solve quantum many-body problems through flexible variational forms \cite{carleo2017_solving}. In
hadron spectroscopy, neural approaches have also been used with Cornell and mixed heavy-quark
potentials \cite{mutuk2019cornell,akan2025predicting}. These studies support the use of learning
methods for nonlinear spectral corrections. Our contribution, however, has a more restricted role.
We do not use ML to replace the potential model. Instead, it is applied after the screened funnel
VMC calculation, so that the physical ordering imposed by the Hamiltonian is preserved and the
learning task is limited to the residual structure.

The broader aim of this work is methodological. Charmonium spectroscopy already has a rich set of
potential-model analyses
\cite{eichten1978charmonium,godfrey1985mesons,barnes2005_higher,ahmad2025charmonium} and precise
experimental anchors for testing numerical predictions \cite{ParticleDataGroup:2024cfk}. This
combination makes it a suitable laboratory for hybrid physics--ML workflows. In our approach, the
screened funnel potential supplies the dynamical baseline, VMC provides the stochastic quantum
solver, and ML supplies the residual map from baseline masses to corrected masses. The paper
evaluates this workflow rather than proposing a new charmonium catalogue. The success of the method
is measured by the improvement over the same VMC baseline, not by claiming that the screened funnel
potential is a definitive description of charmonium. A technical issue is also important in order to
understand the ML phase. In the run under investigation, there are ten charmonium states, of which
seven states use the experimental reference masses and are used in the residual analysis. However,
the ML data set is not limited only to seven points. For each member of charmonium spectroscopy, the
VMC method produces a large set of radial wave functions, local energies, potential values, and
variational parameters. These VMC-generated values are included in the regression process, and then
the results at the sample level are averaged in order to produce one corrected mass for each state.
The following sections define the screened funnel model, describe the VMC and ML procedure, and
quantify the residual correction across the selected states.
\section{Methodology}

We formulate charmonium as a nonrelativistic two-body $c\bar c$ system. This choice follows the
potential-model tradition for heavy quarkonia
\cite{eichten1978charmonium,eichten1980charmonium,godfrey1985mesons}. We therefore use the potential
model as a controlled physics input.

\begin{equation}
H\psi_{nlm}(\mathbf{r})=E_{nl}\psi_{nlm}(\mathbf{r}),
\end{equation}
where $H$ is the relative Hamiltonian, $\psi_{nlm}$ is the charmonium wave function, and $E_{nl}$ is
the binding energy. The quantum numbers $n$, $l$, and $m$ denote the radial, orbital, and magnetic
labels. The reduced mass is defined as
\begin{equation}
\bar m=\frac{m_c}{2},
\end{equation}
since $m_c=m_{\bar c}$ for charmonium. The Hamiltonian is
\begin{equation}
H=-\frac{1}{2 \bar m}\nabla^2+V(r),
\end{equation}
where $r=|\mathbf{r}|$ is the quark--antiquark separation. This form isolates the central
interaction. It also gives a transparent baseline for testing ML residual learning.

The interaction is taken as a screened funnel potential. It combines short-distance Coulomb
attraction with a screened confining term. This structure reflects the physical role of color
Coulomb exchange and confinement in quarkonium models
\cite{eichten1978charmonium,eichten1980charmonium}. Screening is included because heavy-quark bound
states can feel medium-like or channel-induced flattening effects at larger distance
\cite{karsch1988_color,digal2001_quarkonium}. The potential is written as
\begin{equation}
V(r)=-\frac{\kappa}{r}+\frac{\sigma}{\mu_s}\left(1-e^{-\mu_s r}\right)+V_0,
\end{equation}
where $\kappa$ controls the Coulomb strength, $\sigma$ is the string-tension scale, $\mu_s$ is the
screening parameter, and $V_0$ is a constant shift. For the analyzed run, the recorded inputs are
$m_c=1.27~\mathrm{GeV}$, $\sigma=0.273073~\mathrm{GeV}^2$, $\kappa=0.699732$,
$V_0=-0.322296~\mathrm{GeV}$, and $\mu_s=0.128287~\mathrm{GeV}$. In the limit $\mu_s r\ll 1$, the
confining part becomes approximately linear. This connects the model to Cornell-type spectroscopy
\cite{eichten1978charmonium,pathak2022parameterisation,solomko2023cornell}. The expansion is
\begin{equation}
\frac{\sigma}{\mu_s}\left(1-e^{-\mu_s r}\right)=\sigma r-\frac{\sigma\mu_s}{2}r^2+\mathcal{O}(r^3),
\end{equation}
here the first term gives the usual linear confinement. At large $r$, the same term saturates. This
saturation gives a direct mathematical representation of screening.

The radial wave function is separated as
\begin{equation}
\psi_{nlm}(\mathbf{r})=\frac{u_{nl}(r)}{r}Y_{lm}(\theta,\phi),
\end{equation}
where $Y_{lm}$ is the spherical harmonic. The radial equation becomes
\begin{equation}
\left[-\frac{1}{2 \bar m}\frac{d^2}{dr^2}
+\frac{l(l+1)}{2 \bar m r^2}+V(r)\right]u_{nl}(r)=E_{nl}u_{nl}(r),
\end{equation}
where $l(l+1)/(2 \bar m r^2)$ is the centrifugal term. This equation defines the VMC target. Similar
radial reductions are standard in quark--antiquark spectral calculations
\cite{ciftci2003asymptotic,kumar2013asymptotic}. We use them here only to define the numerical
estimator.

The physical mass of a spin-averaged state is computed as
\begin{equation}
M_{nl}^{\mathrm{VMC}}=2m_c+E_{nl},
\end{equation}
$M_{nl}^{\mathrm{VMC}}$ is the raw VMC mass. The term ``VMC baseline'' refers to the physical
baseline with VMC on. The masses that appear in the table are the results of eigensolver
computations and are not averages obtained directly from the Monte Carlo sampling. The
spin-dependent splitting is not fitted separately in this case. Higher charmonium calculations often
include richer spin and transition structures
\cite{barnes2005_higher,deng2017_charmonium,ahmad2025charmonium}. The present design instead tests
whether ML can correct systematic VMC bias within one fixed baseline.

The VMC calculation uses a trial wave function with explicit $n$ and $l$ dependence. We choose a
radial form that enforces regularity at the origin. It also provides exponential decay at large
distance. The trial state is
\begin{equation}
u_{nl}^{T}(r;\boldsymbol{\alpha})=\mathcal{N}_{nl}\,r^{l+1}P_n(r;\boldsymbol{\alpha})e^{-\alpha
r-\beta r^2},
\end{equation}
where $\mathcal{N}_{nl}$ normalizes the state. The vector $\boldsymbol{\alpha}$ contains the
variational parameters. The parameters $\alpha$ and $\beta$ control the radial falloff. The
polynomial $P_n$ supplies radial structure. This ansatz gives sufficient flexibility for low
charmonium levels. It also preserves a compact parameter space for stable VMC optimization. For
excited states, the polynomial factor $P_n(r;\alpha)$ is used to encode the radial-node structure
associated with the chosen $(n,l)$ channel. Thus, each state is treated with a state-specific trial
function rather than by reusing the same radial profile for all levels. The variational parameters
are optimized separately for every $(n,l)$ channel, while the imposed radial form fixes the required
near-origin behavior through $r^{l+1}$ and allows the excited-state structure through $P_n$. This
construction prevents the excited-state calculation from being interpreted as a simple repetition of
the ground-state VMC run. In this work, the VMC energies for the excited channels are therefore
obtained from independently optimized trial states carrying the corresponding radial and orbital
labels.

The probability density sampled in VMC is
\begin{equation}
\rho_{nl}(r;\boldsymbol{\alpha})=\frac{|u_{nl}^{T}(r;\boldsymbol{\alpha})|^2}
{\int_0^\infty |u_{nl}^{T}(r;\boldsymbol{\alpha})|^2dr},
\end{equation}
where the denominator enforces normalization. The local energy is
\begin{equation}
E_L(r;\boldsymbol{\alpha})=\frac{H u_{nl}^{T}(r;\boldsymbol{\alpha})}{u_{nl}^{T}(r;\boldsymbol{\alpha})},
\end{equation}
measures the pointwise action of the Hamiltonian. The variational energy is then
\begin{equation}
E_{nl}^{T}(\boldsymbol{\alpha})=\int_0^\infty \rho_{nl}(r;\boldsymbol{\alpha})E_L(r;\boldsymbol{\alpha})\,dr,
\end{equation}
and the Monte Carlo estimator is
\begin{equation}
\widehat{E}_{nl}^{T}(\boldsymbol{\alpha})=\frac{1}{N_s}\sum_{i=1}^{N_s}E_L(r_i;\boldsymbol{\alpha}),
\end{equation}
where $N_s$ is the number of sampled radial points. The points $r_i$ are distributed according to
$\rho_{nl}$. This estimator is the numerical core of the VMC baseline. The sampled configuration
along with its energy and potentials, which were produced through Monte Carlo sampling process for
each experimental ground state charmonium, are kept as a training instance in ML algorithm, and
hence the dataset is made up of the VMC generated samples rather than seven ground state masses. The
mass value of an experimental ground state is still the common physical anchor point for those
samples corresponding to that state.

The variational parameters minimize the estimated energy. The optimization problem is handled by
\begin{equation}
\boldsymbol{\alpha}_{nl}^{\star}=\arg\min_{\boldsymbol{\alpha}}\widehat{E}_{nl}^{T}(\boldsymbol{\alpha}),
\end{equation}
with $\boldsymbol{\alpha}_{nl}^{\star}$ optimized parameters for each state. This variational
principle connects the method to standard stochastic wave-function optimization. Stochastic
reconfiguration and related Monte Carlo methods show how correlated sampling can stabilize quantum
variational calculations \cite{sorella1998green}. Our implementation uses the same principle of
energy minimization. It keeps the numerical objective explicit and reproducible. In this way, the
screened funnel parameters are optimized at the spectrum level. We define a parameter vector
\begin{equation}
\boldsymbol{\theta}=(m_c,\kappa,\sigma,\mu_s,V_0),
\end{equation}
here each component has a direct physical role. The spectrum-level loss is
\begin{equation}
\mathcal{L}_{\mathrm{VMC}}(\boldsymbol{\theta})=\frac{1}{N_{\mathcal A}}\sum_{j\in\mathcal A}
\left(M_j^{\mathrm{VMC}}(\boldsymbol{\theta})-M_j^{\mathrm{ref}}\right)^2,
\end{equation}
where $\mathcal A$ denotes the experimentally anchored state set and $N_{\mathcal A}=7$ for the
analyzed run. Ten channels are computed in total, while the seven states provide the reference
targets used for parameter calibration and residual learning. The reference values
$M_j^{\mathrm{ref}}$ are the state-level target masses. The loss does not serve as a discovery
criterion. It defines the best numerical resolution of the adopted screened potential.

Neural quantum-state methods show that ML can represent complex many-body structure
\cite{carleo2017_solving}. Neural-network studies of Cornell-type systems also motivate ML as a
numerical surrogate for potential-model errors \cite{mutuk2019cornell,akan2025predicting}. The
network does not replace the Hamiltonian. It learns the remaining mass error after VMC.

For each experimentally verified state $j$, the state-level residual target is
\begin{align}
\Delta_j = M^{\mathrm{ref}}_j - M^{\mathrm{base}}_j,
\qquad M^{\mathrm{base}}_j\equiv M^{\mathrm{VMC}}_j,
\end{align}
where $M^{\mathrm{base}}_j$ denotes the VMC-enabled eigensolver baseline. Although $\Delta_j$ is
anchored by the reference mass of state $j$, it is learned from the VMC sample ensemble generated
for that state. For sampled configuration $i$ in state $j$, the ML input can be represented as

\begin{equation}
x_{ij}=\bigl(r_{ij}, E_L(r_{ij}), V(r_{ij}), n_j, l_j, M^{\mathrm{base}}_j, m_c, \kappa,
 \sigma, \mu_s, V_0, \phi_{ij}\bigr),
\end{equation}
where $\phi_{ij}$ denotes any additional sample-level or variational descriptors retained by the
numerical run. The network predicts a sample-level residual response,
\begin{equation}
\widehat{\Delta}_{ij}=f_w(x_{ij}),
\end{equation}
and the correction assigned to state $j$ is obtained by aggregating its $N_{s,j}$ VMC samples,

\begin{equation}
\widehat{\Delta}_j=\frac{1}{N_{s,j}}\sum_{i=1}^{N_{s,j}}\widehat{\Delta}_{ij}.
\end{equation}
The corrected state mass is therefore
\begin{equation}
M^{\mathrm{ML}}_j=M^{\mathrm{base}}_j+\widehat{\Delta}_j.
\end{equation}
The ML objective is defined over all VMC-generated samples belonging to the anchored state set
$\mathcal A$,

\begin{align}
L_{\mathrm{ML}}(w)=\frac{1}{N_{\mathrm{ML}}}\sum_{j\in\mathcal A}
\sum_{i=1}^{N_{s,j}}\left[\Delta_j-f_w(x_{ij})\right]^2+\eta ||w||_2^2,
\qquad N_{\mathrm{ML}}=\sum_{j\in\mathcal A}N_{s,j}.
\end{align}

Thus, $N_{\mathrm{ML}}$ represents the total number of VMC training samples, which is very large
compared to experimental references. In that stage, VMC ensembles provide the variation at the level
of the sample necessary for the regression. Since all samples from the same state have a common
target to refer to, they may be statistically correlated. Hence, the large number of samples, cannot
be considered as seven independent measurements of the experiment.

As a summary of methodology, the workflow therefore has three linked stages. First, the screened
funnel Hamiltonian creates a physical motivation for the baseline. Next, the optimization and
deterministic eigensolver using the VMC allows the baseline masses to be obtained. Finally, machine
learning learns the residual error on the sample ensembles created using the VMC with the seven
state level references as targets. This is a controlled way of obtaining the final result. The final
result masses are not derived from an unconstrained black box. At the end, we evaluate the residual
correction with compact global metrics, such as MAE, RMSE, and $R^2$. Basically, they measure the
numerical success of the residual correction.

\section{Results and Discussion}

\begin{figure}[htbp]
  \centering
  \includegraphics[width=0.95\linewidth, height=0.85\textheight, keepaspectratio]{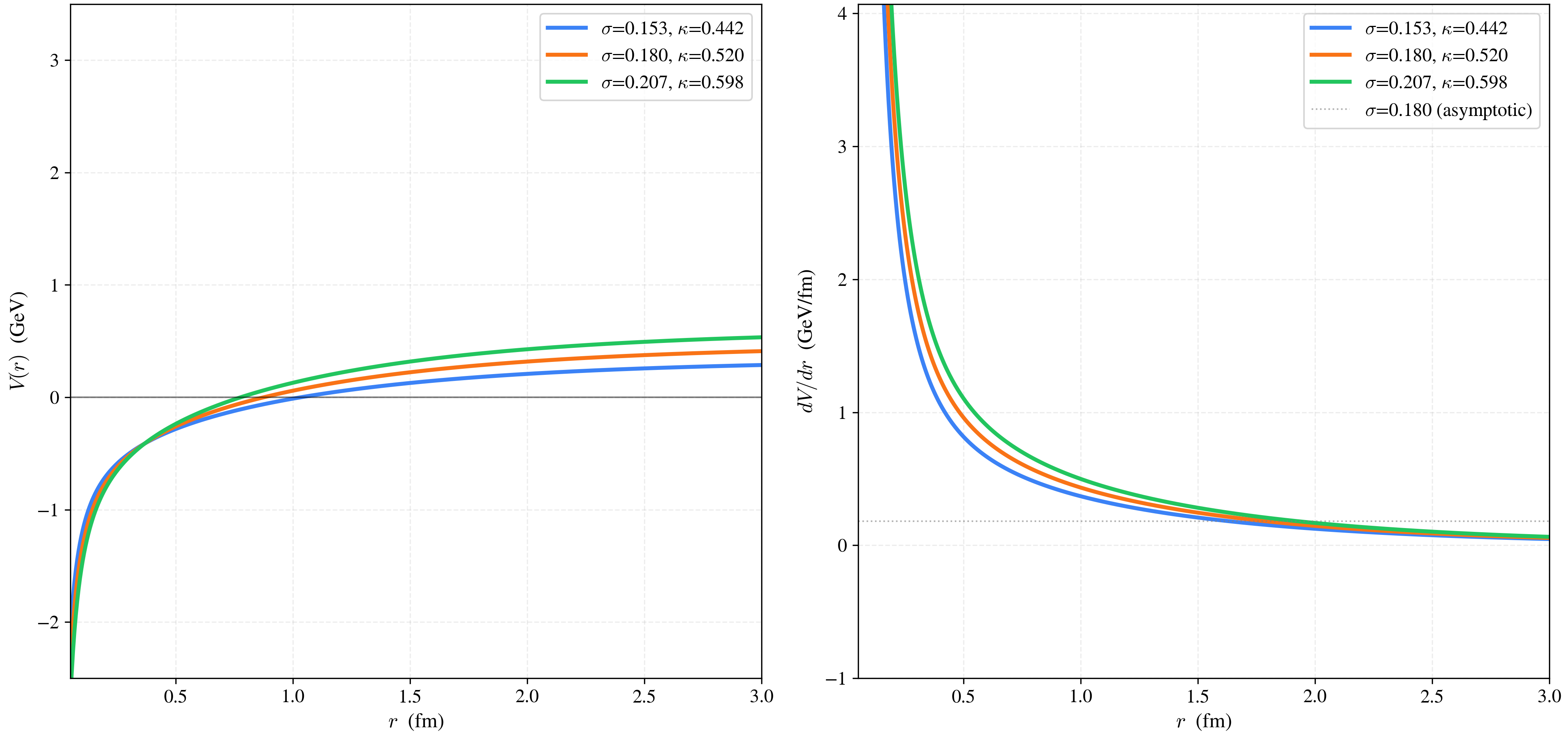}
  \caption{Screened funnel potential and its radial derivative as functions of the interquark
  separation $r$. The left panel shows $V(r)$ for different parameter sets $(\sigma,\kappa)$, while
  the right panel shows $dV/dr$ for the same parameter sets.}
  \label{fig:running_string_tension}
\end{figure}

\begin{figure}[htbp]
  \centering
  \includegraphics[width=0.8\linewidth, height=0.85\textheight, keepaspectratio]{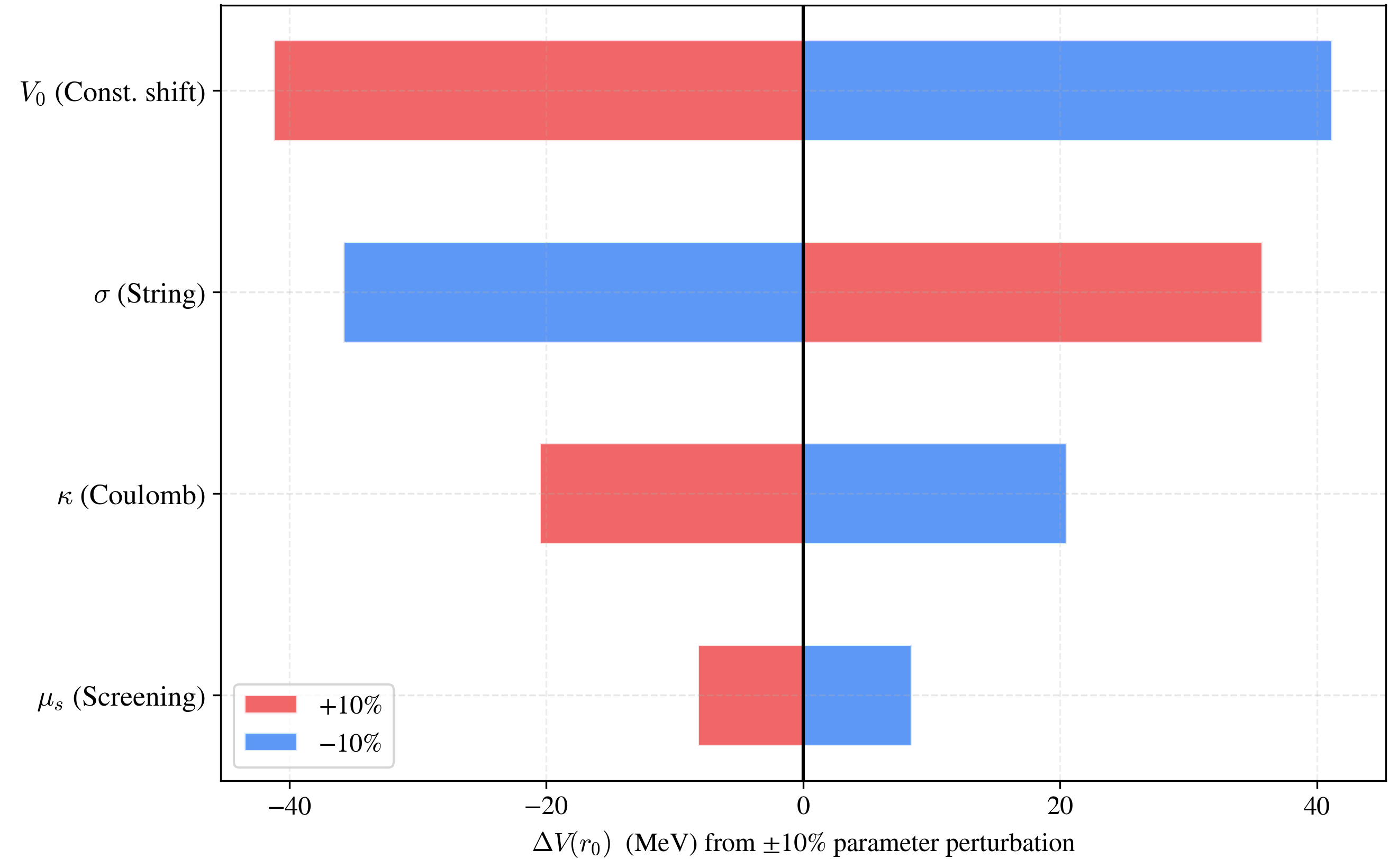}
  \caption{Parameter sensitivity of the screened-funnel baseline. The tornado plot displays the
  change in the potential at the reference radius, $\Delta V(r_0)$, under $\pm 10\%$ variations of
  $V_0$, $\sigma$, $\kappa$, and $\mu_s$.}
  \label{fig:parameter_sensitivity_tornado}
\end{figure}

The numerical results are evaluated as a residual-correction problem applied to the VMC mass
baseline. 
provides the physics baseline. The ML model then learns the residual between this baseline and the
reference charmonium masses. This design keeps the potential model in the calculation. It also
prevents the ML layer from replacing the Schrödinger dynamics with a purely statistical fit. Such a
hybrid strategy matches the present aim. 
classification of the charmonium spectrum. 
many interactions and parameter choices
\cite{eichten1978charmonium,eichten1980charmonium,godfrey1985mesons,pathak2022parameterisation}. The
present test asks a narrower question. It asks whether ML can remove the systematic residuals that
remain after the screened funnel VMC calculation. This question is meaningful because charmonium has
precise reference masses in the Particle Data Group tables
\cite{ParticleDataGroup:2024cfk,patrignani2016review}. It is also meaningful because the screened
interaction contains physics beyond the unscreened Cornell form
\cite{karsch1988_color,digal2001_quarkonium,solomko2023cornell}. The screened funnel potential
controls the baseline before any ML correction is applied. Figure~\ref{fig:running_string_tension}
shows the effective running of the string tension in the screened interaction. The figure
illustrates why the potential differs from a fixed Cornell confinement term. The confinement
strength changes with the interquark distance. This feature is relevant for excited charmonium,
where the wave function samples larger distances than the ground state
\cite{barnes2005_higher,deng2017_charmonium}. Screening therefore modifies the level spacings before
the residual model is trained. On the other hand, Figure~\ref{fig:parameter_sensitivity_tornado}
shows the change in the potential value at the selected reference radius under parameter
perturbations. The dominant entries identify which parameters modify this local potential scale most
strongly This is important for the present argument. The ML correction should improve a tuned
physics baseline. It should not compensate for an uncontrolled parameter choice. Similar parameter
sensitivity appears in Cornell-type studies, where different parameter regions can reproduce
different parts of the heavy-quark spectrum \cite{pathak2022parameterisation,solomko2023cornell}.
The screened funnel form therefore supplies a useful but nontrivial baseline for the VMC
calculation.

\begin{figure}[htbp]
  \centering
  \includegraphics[width=0.95\linewidth, height=0.85\textheight, keepaspectratio]{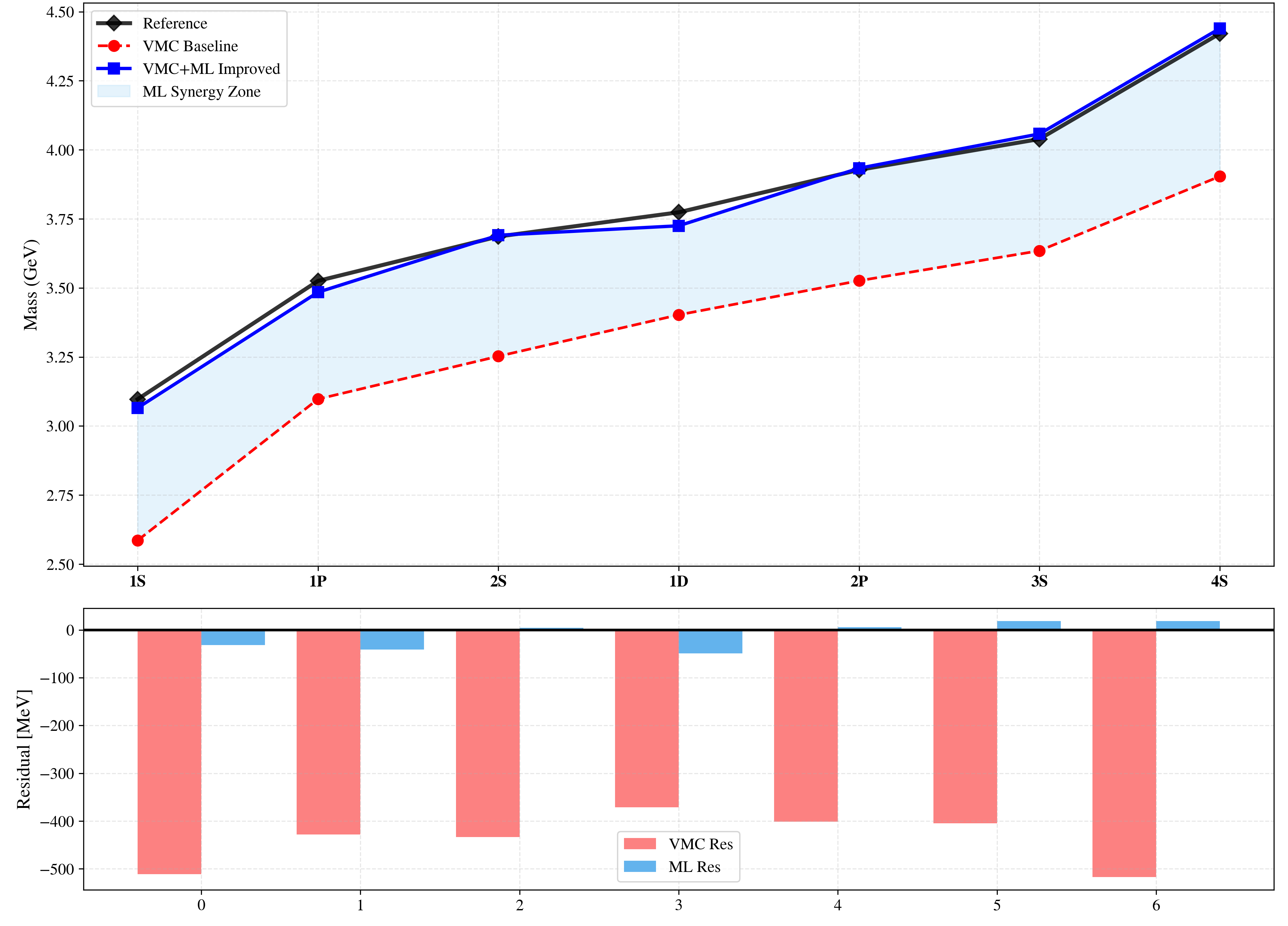}
  \caption{Comparison of the reference targets, raw physics baseline, and ML-corrected charmonium
  masses. The upper panel shows the seven experimentally anchored states from $1S$ to $4S$, while
  the lower panel shows the corresponding baseline and corrected residuals in MeV.}
  \label{fig:procedure_evolution_master}
\end{figure}

The mass and residual comparison is summarized in Fig.~\ref{fig:procedure_evolution_master}. 
for the selected charmonium channels. The lower panel shows the corresponding VMC and ML residuals
in MeV.
The residual correction procedure is then interpreted as an additive correction to the VMC baseline,
where the ML model learns the remaining difference between the VMC prediction and the reference
mass. This procedure differs from direct mass prediction. A direct predictor must learn the full
spectrum from data. A residual predictor learns the missing correction after the physics solver has
already imposed the main structure. This distinction is central to the interpretation of the
results. NN methods have been used to approximate spectra in potential models
\cite{mutuk2019cornell,akan2025predicting}. Neural quantum-state methods also show that ML can
represent quantum many-body structure in variational settings \cite{carleo2017_solving}. The present
implementation uses ML in a more constrained role. It corrects the VMC output, while the screened
Hamiltonian remains the organizing physical input. The relevant ML hyperparameters define this
correction stage. The learning rate controls the update size during training. The number of epochs
controls repeated passes through the training set. The hidden layer size defines the expressivity, 
and the regularization strength constrains overfitting. The train and validation split is done based 
on the sample dataset created by the VMC for the seven channels that are experimentally defined. 
It thus relies on numerous sample-based observations rather than seven observations, while the physical 
performance is assessed afterwards at the level of seven state masses. The correlation of samples within 
the same VMC ensemble, which have the same state name and reference mass, should also be taken into account 
when assessing validation scores and generalization to other channels.

For the seven experimentally verified states, the ML correction reduces the raw-baseline MAE
from $438.13~\mathrm{MeV}$ to $24.12~\mathrm{MeV}$, corresponding to a $94.5\%$ improvement.
This reduction measures the effect of residual learning at the sample level on the 
VMC-generated ensembles after they are grouped into seven states. It is not a direct refit of seven 
isolated mass points. The corrected spectrum gives $\mathrm{RMSE}=28.77~\mathrm{MeV}$,
$R^2=0.9944$, $\mathrm{MAPE}=0.66\%$, a maximum absolute error of $48.82~\mathrm{MeV}$ for the
$1D$ state, and a mean signed error of $-10.53~\mathrm{MeV}$. The remaining three channels, $2D$,
$3P$, and $3D$, are predictions and are excluded from these experimental metrics.

\begin{table}[htbp]
  \centering
  \caption{The mass comparison for selected charmonium radial-orbital channels.}
  \label{tab:prediction_error_table}
  \small
  \begin{adjustbox}{max width=\linewidth}
  \begin{tabular}{l c c c c c c}
  \toprule
  State
  & Experimental \cite{ParticleDataGroup:2024cfk}
  & ML corrected
  & Raw baseline
  & \cite{M2025Mass}
  & \cite{barnes2005_higher} (NR)
  & \cite{deng2017_charmonium} (SP) \\
  \midrule

  1S & 3.09693 & 3.06548 & 2.5857 & 2.976 & 2.982 & 2.984 \\

  1P & 3.52538 & 3.48440 & 3.0974 & - & 3.516 & 3.526 \\

  1D & 3.77380 & 3.72498 & 3.4026 & 3.705 & 3.785 & 3.792 \\

  2S & 3.68610 & 3.69072 & 3.2525 & 3.550 & 3.630 & 3.637 \\

  2P & 3.92740 & 3.93339 & 3.5260 & - & 3.934 & 3.916 \\

  2D & - & 4.08914 & 3.7259 & 4.158 & 4.142 & 4.095 \\

  3S & 4.03900 & 4.05762 & 3.6346 & 4.111 & 4.072 & 4.030 \\

  3P & - & 4.24602 & 3.8170 & - & 4.279 & 4.193 \\

  3D & - & 4.43561 & 3.9717 & - & - & 4.336 \\

  4S & 4.42100 & 4.43934 & 3.9039 & 4.523 & 4.406 & 4.281 \\

  \bottomrule
  \end{tabular}
  \end{adjustbox}

  \vspace{2mm}
\end{table}

There are some different spectral notations in the literature. The seven adopted experimental
states in the run are $J/\psi(1\,{}^3S_1)$ for $1S$, $h_c(1\,{}^1P_1)$ for $1P$,
$\psi(3770)$ for $1D$, $\psi(2S)$ for $2S$, $\chi_{c2}(3930)$ for $2P$, $\psi(4040)$ for $3S$, and
$\psi(4415)$ for $4S$. No experimental value is used for $2D$, $3P$, or $3D$.

\begin{figure}[htbp]
  \centering
  \includegraphics[width=0.8\linewidth, height=0.85\textheight, keepaspectratio]{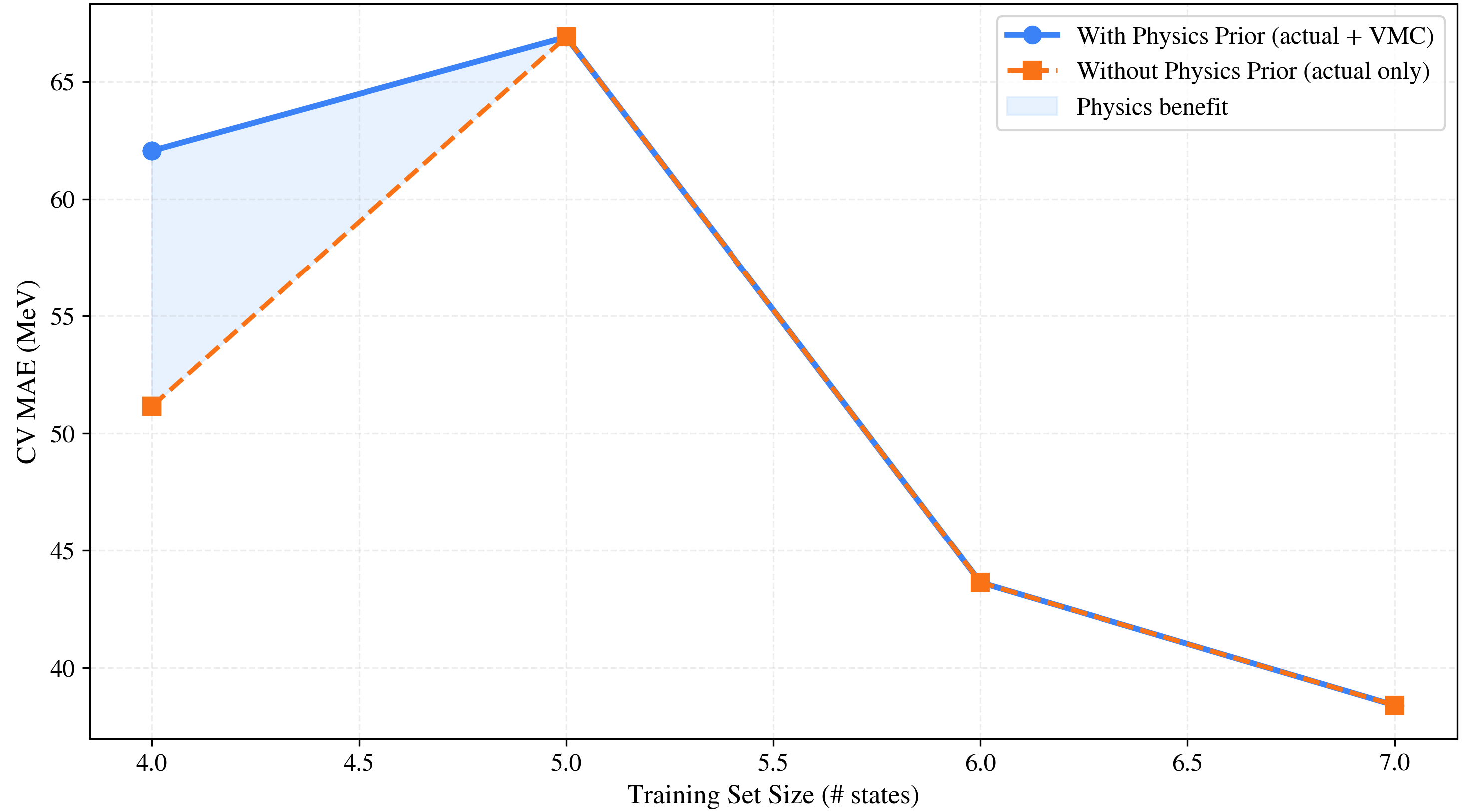}
  \caption{Cross-validation mean absolute error as a function of training-set size. The plot
  compares residual learning with a physics prior, using actual and VMC-based inputs, against
  learning without the physics prior, using actual inputs only. The stated training-set size
  counts included state groups and each group contributes its corresponding VMC sample ensemble.}
  \label{fig:learning_curve_physics}
\end{figure}

\begin{figure}[htbp]
  \centering
  \includegraphics[width=0.8\linewidth, height=0.85\textheight, keepaspectratio]{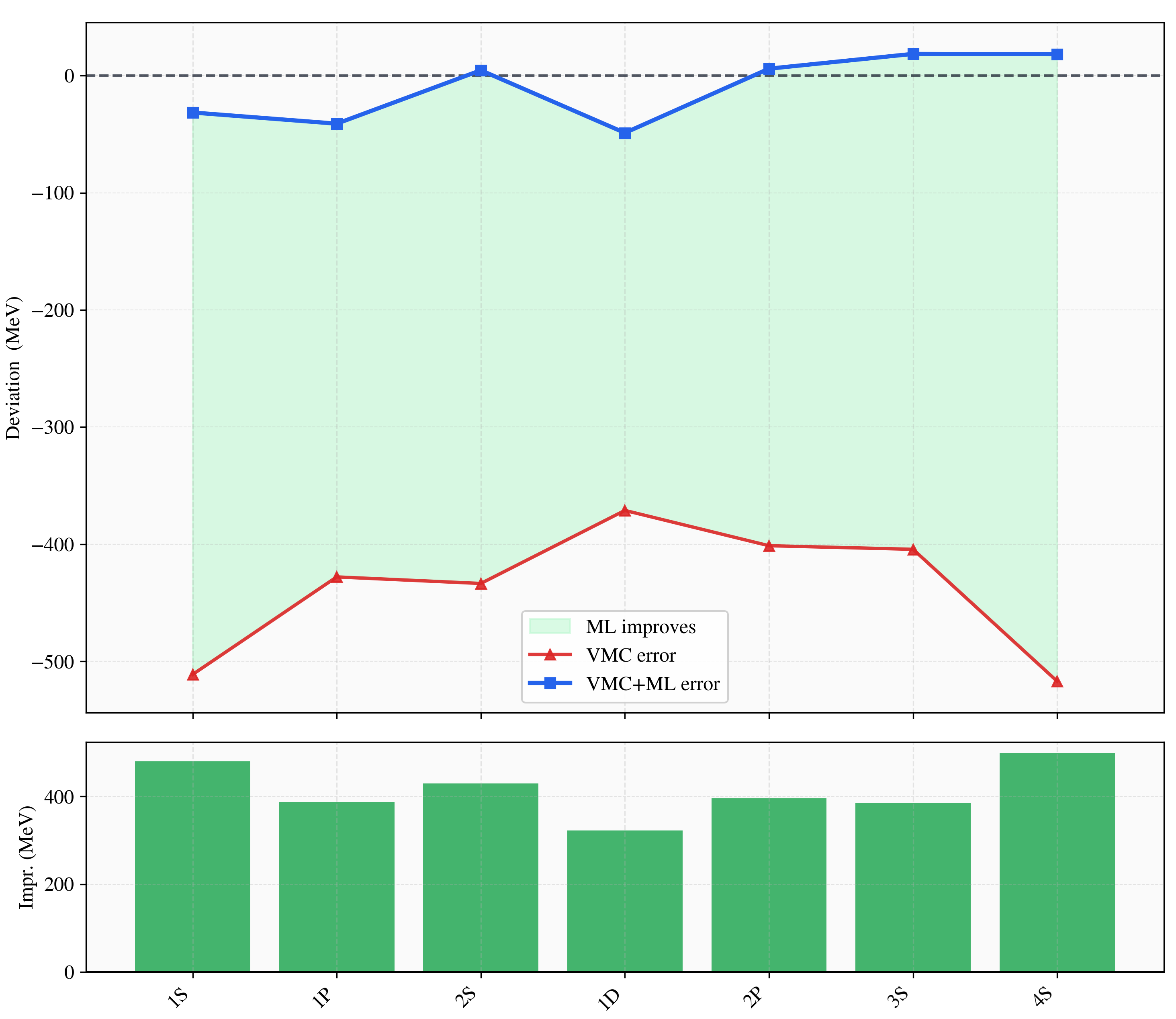}
  \caption{Residual and improvement comparison for the seven experimentally anchored charmonium
  states. The upper panel shows raw-baseline and ML-corrected deviations from the adopted
  references, while the lower panel gives the absolute improvement for each state.}
  \label{fig:comparison_error_bands}
\end{figure}

Figure~\ref{fig:learning_curve_physics} illustrates the impact of the training-set size on the
residual model. The plotted value is the cross-validation MAE for learning both with and without
the physics-prior information. For four included state groups, the curve with the physics prior
shows the greater MAE, whereas the curves coincide starting from five to seven groups. However,
each data point still consists of the VMC sample ensembles for the included states, so that the
horizontal axis reflects the number of spectroscopic groups instead of the number of ML samples.
The presented graph does not indicate any obvious advantage of the validation of the prior in
this case but provides insight into the effect of introducing additional physical state groups to
the regression. This condition is more stringent than the fitting of the training points. It checks
whether the model learns the smooth residual pattern among the charmonium states.
Figure~\ref{fig:comparison_error_bands} presents the same result in the target space employed in
the analyzed ML run. The corrected masses differ from the adopted references by no more than
$49~\mathrm{MeV}$, while the uncorrected masses exhibit systematic negative deviations by
approximately $370$--$520~\mathrm{MeV}$. The depicted values correspond to the same seven-state
errors used to determine the global metrics provided in
Table~\ref{tab:prediction_error_table}. The model corrects a dominant structured offset rather than
small fluctuations. Such residual correction is physically reasonable due to the fact that the
potential models usually learn the level ordering better than their absolute locations.

\begin{figure}[htbp]
  \centering
  \includegraphics[width=0.95\linewidth, height=0.85\textheight, keepaspectratio]{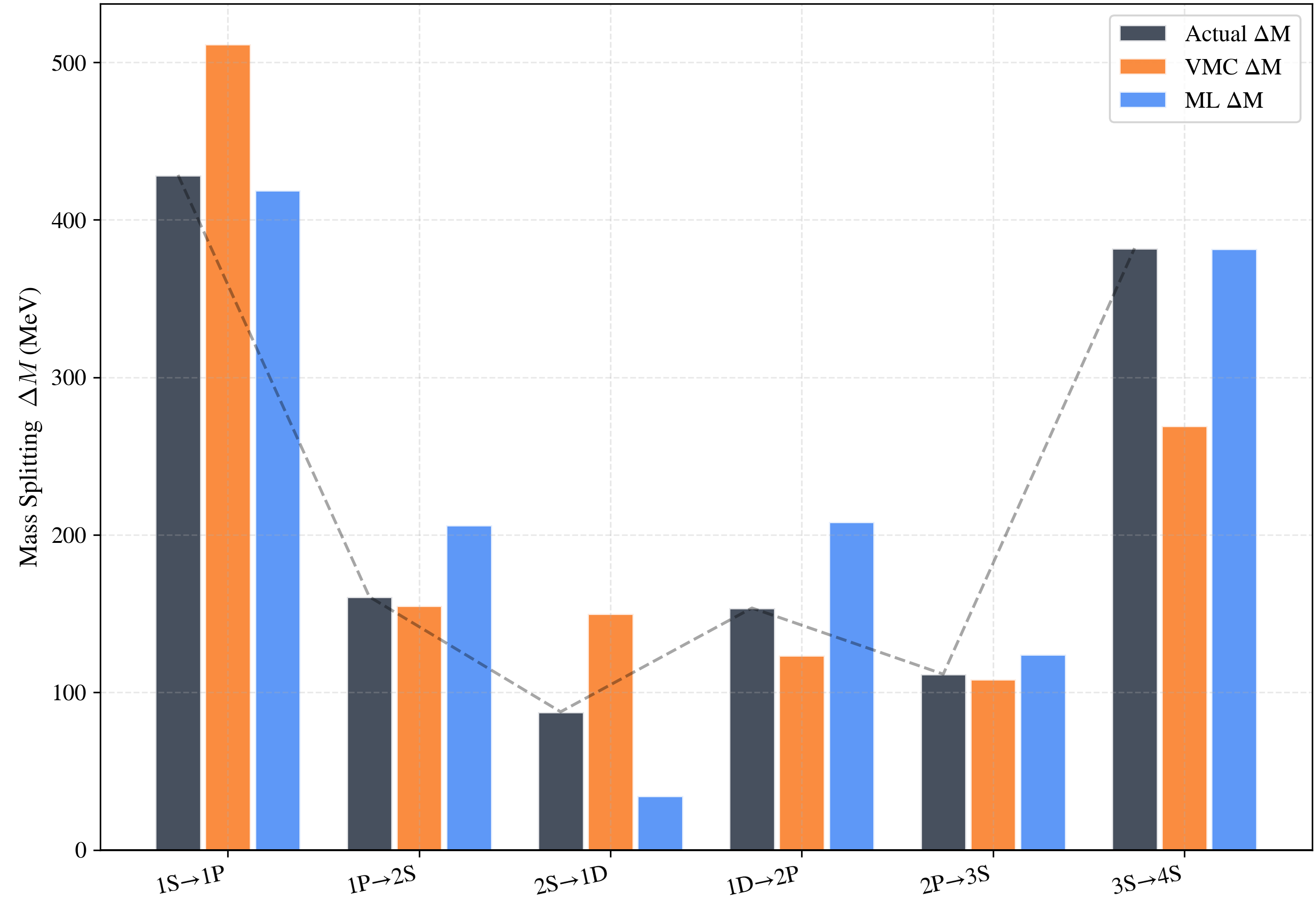}
  \caption{Comparison of charmonium mass splittings for consecutive anchored-state pairs. The
  grouped bars show the adopted-reference, raw-baseline, and ML-corrected splittings for the
  transitions from $1S\!\to\!1P$ through $3S\!\to\!4S$.}
  \label{fig:mass_splittings}
\end{figure}

Moreover, figure~\ref{fig:mass_splittings} adds a more physical view. It compares the splittings
between neighboring or related charmonium levels. Mass splittings test the internal structure of the
spectrum. They are less sensitive to a common offset than absolute masses. The figure therefore
probes whether the ML correction preserves the spectroscopy imposed by the screened VMC baseline.
The global ordering is preserved, but the correction does not improve every individual splitting: 
the $1P\!\to\!2S$ and $1D\!\to\!2P$ gaps are overestimated, while the $2S\!\to\!1D$ gap is underestimated. 

\begin{table}[htbp]
  \centering
  \caption{Test set MAE and RMSE values for the regression models used in the residual-correction
  stage.}
  \label{tab:summary_metrics}
  \small
  \begin{adjustbox}{max width=\linewidth}
  \begin{tabular}{l c c}
    \toprule
    Method & Test MAE (GeV) & Test RMSE (GeV) \\
    \midrule
    ExtraTrees & $0.068974\pm0.000311$ & $0.084481\pm0.000521$ \\
    GradientBoosting & $0.039944\pm0.000532$ & $0.048916\pm0.000716$ \\
    KNN & $0.013841\pm0.000681$ & $0.018235\pm0.000894$ \\
    MLP & $0.002334\pm0.000291$ & $0.003178\pm0.000351$ \\
    RandomForest & $0.071108\pm0.000777$ & $0.088543\pm0.001015$ \\
    Ridge & $0.029055\pm0.000212$ & $0.039539\pm0.000153$ \\
    SVR & $0.003961\pm0.000052$ & $0.004903\pm0.000044$ \\
    \bottomrule
  \end{tabular}
  \end{adjustbox}
\end{table}

\begin{figure}[htbp]
  \centering
  \includegraphics[width=0.95\linewidth, height=0.85\textheight, keepaspectratio]{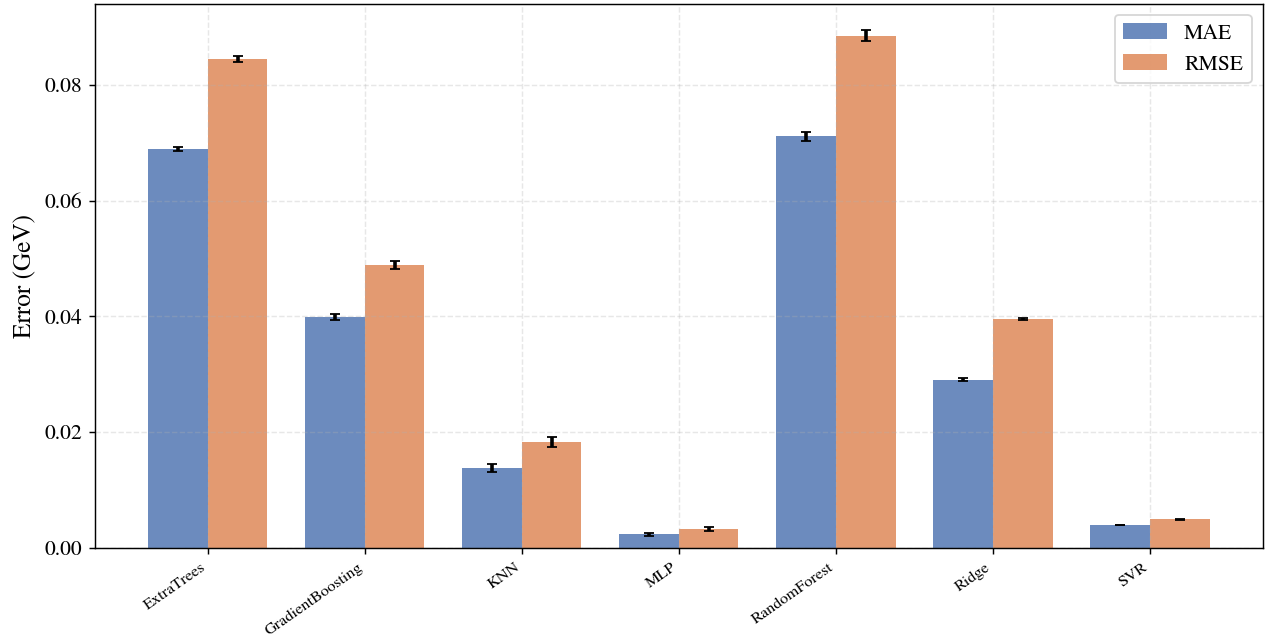}
  \caption{Test-set MAE and RMSE values for the regression models used in the residual-correction
  stage.}
  \label{fig:mae_rmse_bar}
\end{figure}

MLP gives the smallest test MAE and RMSE, followed by SVR and KNN. RandomForest gives the
largest errors in the updated model comparison. These seed-averaged values describe regression over
the VMC-generated sample ensembles; they are distinct from the seven aggregated state-mass errors
reported in Table~\ref{tab:prediction_error_table}. Figure~\ref{fig:mae_rmse_bar} compares the test set MAE and RMSE values listed in Table~\ref{tab:summary_metrics} 
and stands behind the result that the MLP gives the smallest errors, followed by SVR, whereas RandomForest 
gives the largest values.

\begin{figure}[htbp]
  \centering
  \includegraphics[width=0.8\linewidth, height=0.85\textheight, keepaspectratio]{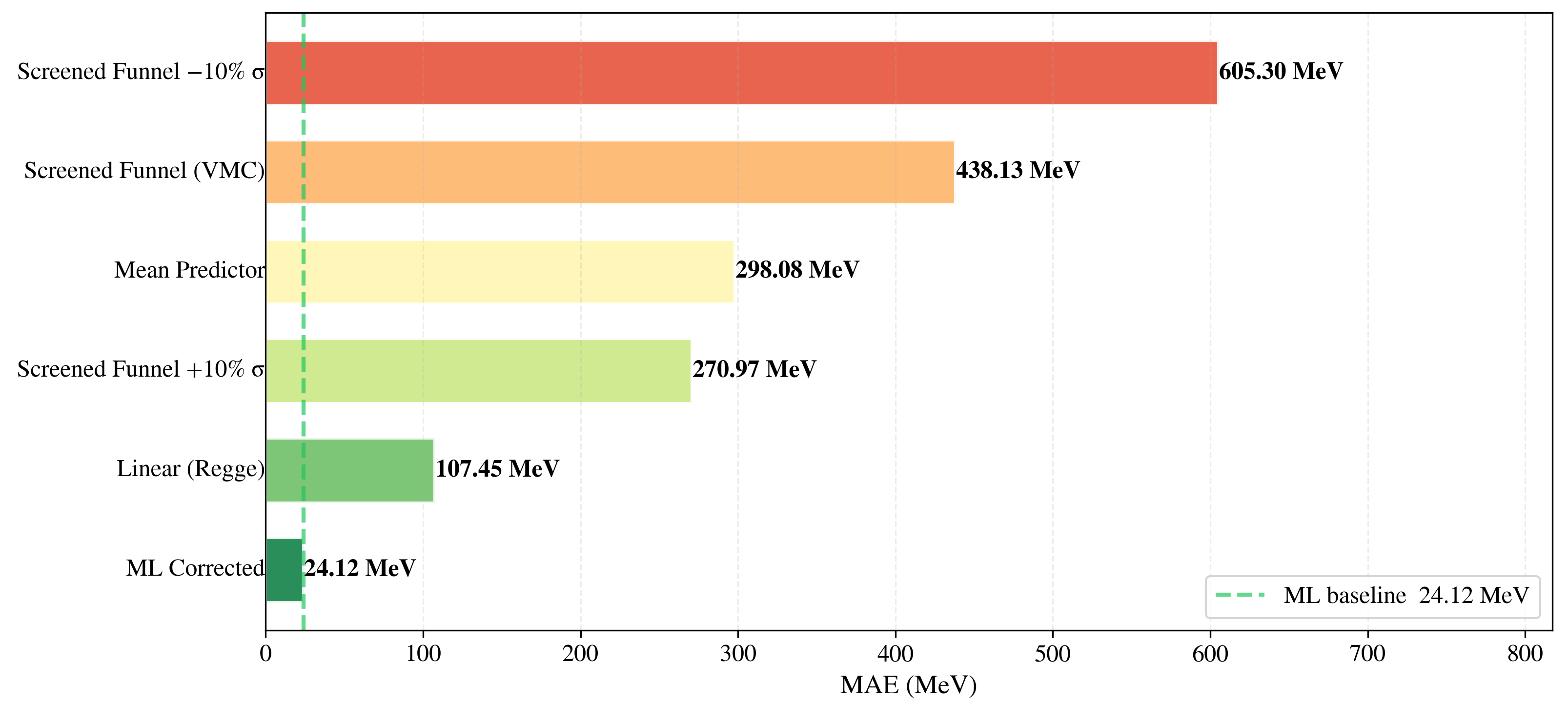}
  \caption{MAE comparison of the baseline and correction diagnostics generated for the current
  seven-state anchored evaluation. The bars list the mean predictor, $\pm10\%$ string-tension
  variations, the raw screened-funnel baseline, the Ridge diagnostic, and the ML-corrected result in
  MeV.}
  \label{fig:potential_model_tournament}
\end{figure}

Figure~\ref{fig:potential_model_tournament} places the screened funnel baseline and correction
results beside the diagnostic alternatives generated in the current run. For the experimentally verified states, the raw screened-funnel baseline has an MAE of $438.13~\mathrm{MeV}$ and the
residually corrected result has an MAE of $24.12~\mathrm{MeV}$. The figure also reports
$605.30~\mathrm{MeV}$ for the $-10\%$ string-tension perturbation, $270.97~\mathrm{MeV}$ for the
$+10\%$ perturbation, $298.08~\mathrm{MeV}$ for the mean predictor, and $107.45~\mathrm{MeV}$ for
the linear Ridge diagnostic. The comparison is intended to assess the behavior of the present
hybrid calculation rather than to establish the universal superiority of a particular potential.
Existing Cornell, relativized, exponential, and screened descriptions differ in their treatments of
confinement, short-range dynamics, spin effects, and screening
\cite{eichten1978charmonium,eichten1980charmonium,godfrey1985mesons,sreelakshmi2022mass,ahmad2025charmonium,M2025Mass}.
The present comparison has a more specific purpose: it shows that the final result should be
interpreted as the coupled physics--ML procedure rather than the baseline alone. This result also
extends earlier neural-network applications to potential-model spectroscopy
\cite{mutuk2019cornell,akan2025predicting}. Here, however, the learned correction is applied to the
output of the VMC-enabled potential-model workflow. This formulation is compatible with broader
stochastic and variational approaches to quantum systems \cite{sorella1998green,carleo2017_solving}.

Placing the present calculation beside earlier charmonium studies helps define its physical scope.
Remember that conventional potential models organize the spectrum and provide useful dynamical
intuition \cite{eichten1994mesons,godfrey1985mesons}, while screened interactions are particularly
relevant for higher states that probe larger interquark separations
\cite{karsch1988_color,digal2001_quarkonium,solomko2023cornell}. Recent studies further refine this
description through alternative interactions and transition operators
\cite{deng2017_charmonium,ahmad2025charmonium}. Accordingly, the aim is not to introduce new
quantum-number assignments, but to use the established charmonium spectrum as a controlled test of
whether residual learning can improve the physics baseline. The learning-curve and mass-splitting 
diagnostics show that the global offset is strongly reduced and the ordering is retained. The ML 
training set itself contains many VMC samples for each state, and the main remaining limitation is 
the finite number of distinct state categories available for testing state-to-state transfer, together 
with several distorted individual splittings.

The residual model has limits that follow from the data and design. The analyzed ML run
contains a large VMC-generated sample ensemble for each of the seven experimentally anchored
channels, rather than only seven training observations. This sample-level richness supports
nonlinear regression and explains why the MLP can be trained meaningfully. Nevertheless, samples
within the same channel share a common physical target and are not equivalent to independent new
spectroscopic states. A highly flexible model could exploit within-state redundancy without
necessarily transferring to a new radial-orbital channel. Although MLP gives the lowest
seed-averaged regression errors in Table~\ref{tab:summary_metrics}, transferability to the
unobserved $2D$, $3P$, and $3D$ channels should be tested with state-held-out or independently
generated ensemble validation. Lower-variance linear or regularized models should therefore remain
part of future comparisons.

A more fundamental limitation is the magnitude of the baseline offset. The run uses
$m_c=1.27~\mathrm{GeV}$, corresponding to the commonly quoted
$\overline{\mathrm{MS}}$ charm-mass scale, whereas nonrelativistic potential
calculations generally require an effective or pole-like constituent mass. The resulting
$438~\mathrm{MeV}$ baseline MAE and the nearly uniform
$400$--$500~\mathrm{MeV}$ corrections indicate that the neural network is largely
learning an additive mass renormalization. The fitted values
$\sigma=0.273073~\mathrm{GeV}^2$ and $\kappa=0.699732$ are larger than the
values commonly employed in Cornell-type and related charmonium potential models
\cite{eichten1978charmonium,eichten1980charmonium,godfrey1985mesons,
barnes2005_higher,pathak2022parameterisation} and may therefore be compensating
for the low mass input. By contrast, the screening parameter
$\mu_s=0.128287~\mathrm{GeV}$ remains within a physically reasonable range and
exhibits weaker local sensitivity. Nevertheless, because all potential parameters are
determined simultaneously, $\mu_s$ may still be correlated with $m_c$, $V_0$,
$\sigma$, and $\kappa$ through the fitting procedure. A physically interpretable
follow-up should first repeat the calculation with a pole-like charm mass near
$1.5~\mathrm{GeV}$ and then refit all screened-funnel parameters before applying ML.

Still, the corrected output should be interpreted as a sample-level
residual-corrected result within the chosen screened funnel potential. It should not be interpreted as an independent solution of QCD. Lattice
and effective nonrelativistic QCD calculations address heavy-quark systems from different starting
points \cite{assi2023baryons,assi2024tetraquarkssufficientlyunequalmassheavy}. Potential models
instead provide controlled phenomenological Hamiltonians for spectroscopy
\cite{eichten1978charmonium,eichten1980charmonium}. The present contribution belongs to the second
category. The methodological contribution is therefore the coupling of the VMC-enabled physics
baseline with residual ML correction. Numerically, this coupling lowers the baseline mass error,
while physically it retains the spectral organization generated by the screened Hamiltonian.
\section{Conclusion}
This study examined whether ML residual correction can improve a VMC-enabled eigensolver
baseline for the charmonium spectrum obtained with a screened funnel interaction. The potential
supplies the short-distance Coulomb attraction and screened long-distance confinement, while the
combined variational and diagonalization workflow provides the physical baseline.

The ML stage learns the residual difference between the baseline and the targets
from the VMC-generated sample ensembles of the seven experimental values. The training data therefore
contain many samples for each state, and the learned responses are aggregated to
obtain the final corrected masses. It therefore supplements rather than replaces the
Hamiltonian-based calculation. The corrected spectrum retains the level ordering generated by the
screened-potential baseline. For the seven experimentally anchored states, the MAE decreases
from $438.13~\mathrm{MeV}$ to $24.12~\mathrm{MeV}$, corresponding to a $94.5\%$ reduction, with
$\mathrm{RMSE}=28.77~\mathrm{MeV}$, $R^2=0.9944$, and $\mathrm{MAPE}=0.66\%$. The largest remaining
deviation is $48.82~\mathrm{MeV}$ for the $1D$ state.

Overall, the results demonstrate that residual learning can reduce systematic mass errors
while preserving the main level ordering supplied by the screened funnel potential. Despite this, however, 
the ML correction itself possesses a considerable, roughly scheme-independent contribution. Such 
an effect can be, to some extent, explained by the correlations between the chosen value of the 
charm-quark mass, the shift of the potential due to the constant term, and the rest of the screened 
funnel parameters. As all the parameters used here have been determined within the same framework, 
they are to be treated as effective parameters of the current approach, rather than a scheme-independent 
determination. The focus of the current investigation is the enhancement of the prediction obtained 
from the combination of the VMC data with ML. The remaining deviations in several excited channels 
indicate that the method should be interpreted as a VMC-based
residual improvement of the potential model rather than an independent solution of QCD. The same workflow 
can be extended to other heavy-quark systems in which a physically motivated result exhibits structured 
numerical residuals. The remaining deviations in several excited channels 
indicate that the method should be interpreted as a VMC-based
residual improvement of the potential model rather than an independent solution of QCD. The same workflow 
can be extended to other heavy-quark systems in which a physically motivated baseline exhibits structured 
numerical residuals \cite{monteiro2017cb,akan2025spin,assi2023baryons,assi2024tetraquarkssufficientlyunequalmassheavy}.

\section*{Acknowledgements}

This research was supported by the Scientific Research Projects Coordination Unit of Yozgat Bozok
University (BAP) under Project No.~FBG-2025-1799. The authors also acknowledge YEKUT, the High
Energy and Quantum Computations Laboratory at Yozgat Bozok University, where the computational part
of this study was carried out.

\bibliographystyle{unsrt}
\bibliography{references}
\end{document}